\documentclass[10pt,numbers]{sigplanconf}

\usepackage{acmcopyright}
\usepackage{amsmath}

\usepackage{color}
\usepackage{graphicx}
\usepackage[normalem]{ulem}
\usepackage{listings}
\usepackage{pgf}
\usepackage{tikz}
\usetikzlibrary{positioning}
\usepackage{subfig}
\usepackage{url}
\usepackage{soul}
\usepackage[hidelinks]{hyperref}

\usetikzlibrary{decorations.pathreplacing}
\usetikzlibrary{matrix}
\tikzset{table matrix/.style={draw=black,thick,inner sep=0,matrix of nodes, nodes in empty cells,%
        nodes={minimum width=30mm,minimum height=5mm,draw,outer sep=0,inner sep=0},
              }
          }

\newcommand{\circled}[1]{\raisebox{.4pt}{\textcircled{\raisebox{-1.0pt}{#1}}}}

\begin{document}

\setlength{\pdfpageheight}{\paperheight}
\setlength{\pdfpagewidth}{\paperwidth}

\CopyrightYear{2017}
\setcopyright{acmcopyright}
\conferenceinfo{VEE '17,}{April 08-09, 2017, Xi'an, China}
\copyrightdata{978-1-4503-4948-2/17/04}
\copyrightdoi{http://dx.doi.org/10.1145/3050748.3050763}




\title{Security Analysis of Encrypted Virtual Machines}


\authorinfo{Felicitas Hetzelt}
           {Technical University of Berlin\\Berlin, Germany}
           {file@sec.t-labs.tu-berlin.de}
\authorinfo{Robert Buhren}
           {Technical University of Berlin\\Berlin, Germany}
           {robert@sec.t-labs.tu-berlin.de}

\maketitle

\begin{abstract}

Cloud computing has become indispensable in today's computer landscape.
The flexibility it offers for customers as well as for providers has become a crucial factor for large parts of the computer industry.
Virtualization is the key technology that allows for sharing of hardware resources among different customers.
The controlling software component, called hypervisor, provides a virtualized view of the computer resources and ensures separation of different guest virtual machines.
However, this important cornerstone of cloud computing is not necessarily trustworthy or bug-free.
To mitigate this threat AMD introduced Secure Encrypted Virtualization, short SEV, 
which transparently encrypts a virtual machines memory.

In this paper we analyse to what extend the proposed features can resist a malicious hypervisor 
and discuss the trade-offs imposed by additional protection mechanisms.
To do so, we developed a model of SEV's security capabilities based on the
available documentation as
actual silicon implementations are not yet on the market. 

We found that the first proposed version of SEV is not up to the
task owing to three design shortcomings. 
First the virtual machine control block is not encrypted and handled directly
by the hypervisor, allowing it to bypass VM memory encryption by
executing conveniently chosen gadgets. 
Secondly, the general purpose registers are not encrypted upon \textit{vmexit}, 
leaking potentially sensitive data.
Finally, the control over the nested
pagetables allows a malicious hypervisor to
closely monitor the execution state of a VM and attack it with memory replay attacks.
\end{abstract}




%



%
%
\section{Introduction}
\label{sec:introduction}
Cloud computing has been one of the most prevalent trends in the computer industry in the last decade.
It offers clear advantages for both customers and providers.
Customers can easily deploy multiple servers and dynamically allocate resources according to their immediate needs.
Providers can ver-commit their hardware and thus increase the overall utilization of their systems.
The key technology that made this possible is virtualization, it allows multiple operating systems to share hardware resources.
The hypervisor is responsible for providing temporal and spatial separation of the virtual machines (VMs).
However, besides these advantages virtualization also introduced new risks.

Customers who want to utilize the infrastructure of a cloud provider must fully trust the cloud provider.
Especially the hypervisor is a critical component provided by the cloud hoster as it has full control over the guest VMs.
A malicious or compromised hypervisor can read and write the entire guest memory. 
This affects the integrity and confidentiality of the customer's secrets and the integrity of the customer's services.
Security issues might lead to a full breach of the hypervisor through a hosted VM; 
bugs of such severity have been reported for all most commonly used hypervisors~\cite{cveVirtualBox, cveXen, cveVmware, cveHyperV, cveVenom}.
As a single cloud instance often hosts multiple guest VMs from different customers such security issues allow a malicious tenant to steal confidential data from other customers.

Intel's Software Guard Extensions (SGX)~\cite{sgx2016} and AMD's Secure Encrypted Virtualization (SEV)~\cite{sme2016} are industry's answer to these threats.
They extend the features of the processor to reduce the impact of a malicious, higher privileged software in regards to the confidentiality and integrity of lower privileged software.
SGX enables the customer to create a secure enclave where special code can be executed in a trusted environment that cannot be tampered with by the hypervisor or the operating system.
SGX achieves this by requiring the customer to identify the security sensitive parts of a program and to alter them such that these parts are executed in an SGX enclave.
SEV, on the other hand, allows a customer to encrypt the unaltered VM's memory so that the hypervisor is not able to inspect its data.
The recent addition of SEV Encrypted State (SEV-ES) extends the cryptographic protection of the guest VM to its control state and its general purpose registers.
As can be seen from the AMD SEV whitepaper~\cite{sme2016}:
\begin{quote}
    \textit{,,SEV technology is built around a threat model where an attacker is assumed to have access to not only execute user level privileged code on the target machine, but can potentially execute malware at the higher privileged hypervisor level as well.  The attacker may also have physical access to the machine including to the DRAM chips themselves.  In all these cases, SEV provides additional assurances to help protect the guest virtual machine code and data from the attacker.''}
\end{quote}
The advantage of a solution such as AMD's SEV is that it can be easily adopted by customers because no changes to their existing application software are needed.

While the research community has examined Intel's SGX \cite{costanintel, Weichbrodt, Xu2015}, AMD's SEV has not been subject to scientific research so far.
It is thus unclear what level of protection SEV can provide.
In this paper, we have a first look at the upcoming AMD SEV technology based on publicly available documentation.
We identify possible design issues that can be leveraged by a malicious hypervisor to compromise the guest VM.
To that end, we implement in total three proof of concept attacks on a currently available system.
For the construction of the attacks, 
we bear in mind not only the restrictions an AMD SEV-enabled system imposes,
but also evaluate how the initial SEV design 
could be hardened without sacrificing further guest transparency or impacting cloud maintenance operation.
However we show that even an attacker restricted to basic resource management capabilities,
is still able to gain access to the protected guest system.\\
Our contributions are:
\begin{itemize}
    \item[-] We show how a malicious hypervisor can coerce the guest to
        leak arbitrary memory content and perform arbitrary write operations on encrypted memory.
    \item[-] We describe how to completely disable any memory encryption configured by the tenant.
    \item[-] We implement a replay attack that uses captured login data to gain access to the target system by solely exploiting resource management features of a hypervisor.
\end{itemize}
For the first two attacks we base our evaluation on the initial design of SEV without the optional SEV-ES extension.
The third attack considers the protection of the guest state and general purpose register, as it might be implemented by SEV-ES.
As processors featuring SEV are not available yet, 
it is unclear whether our results will apply to real silicon implementations or future versions of SEV.
We therefore emphasize that we did not break AMD SEV itself but rather evaluated the 
design issues present in the documentation with respect to their capability to protect a guest VM against a malicious or compromised hypervisor.

The rest of the paper is structured as follows: In Section~\ref{sec:background} we give an overview on x86 virtualization and AMD SEV.
We evaluate the security of the protection mechanisms proposed by SEV in Section~\ref{sec:sev_considerations} and discuss our attack model in Section~\ref{sec:attack_model}.
In Section~\ref{sec:attacks} we present our attack.
We discuss possible mitigations to our attack in Section~\ref{sec:discussion}.
In Section~\ref{sec:related_work} we evaluate alternative approaches to shield execution environments from higher privileged adversaries
and present related attacks under similar threat models.
Finally, we discuss future work in Section~\ref{sec:future_work} and conclude our work in Section~\ref{sec:conclusion}.

\section{Background}
\label{sec:background}
In this section we first give a brief introduction to x86 virtualization then we discuss the design of AMD's SEV technology.
This information by no means represents a complete overview of these topics.
The specification for AMD SVM and AMD SEV are however publicly available.
Thus, we refer the interested reader to~\cite{amd64architecture, sev2016}.

\subsection{x86 Virtualization Technologies}
\label{subsec:x86_virt}
In 2005, both Intel (VT-x) \cite{uhlig2005} and AMD (SVM) \cite{svm2005} introduced hardware extensions to their x86 processors that added a higher privileged mode to the existing ring 0 to ring 3 privilege levels.
This new mode, called host mode, comprises another set of the privilege rings 0 to 3 and is higher privileged than the non-host mode, called guest mode.
The host mode is intended to host the hypervisor whereas the guest VM usually executes in the non-host mode.
To make use of these extensions, a hypervisor, running in the host mode, uses a special instruction, \texttt{vmrun}, to switch the CPU to the guest mode.
This instruction takes the address of a control structure as a single argument in the register \texttt{rAX}.
This control structure, called \texttt{vmcb}, contains the guest state, entry controls (pending virtual interrupts) and exit controls.
Prior to the initial start of the guest, the hypervisor configures the \texttt{vmcb} and initializes the general purpose registers as they are not part of the \texttt{vmcb}.
Upon issuing the \texttt{vmrun} instruction, the CPU copies the values of the \texttt{vmcb} fields into the respective hardware registers and starts execution of the guest at the entry point defined in the \texttt{vmcb}.
An event that is flagged in the \texttt{vmcb} as such will lead to a \texttt{vmexit} with the exit reason set in the \texttt{vmcb}.
The hypervisor then handles the exit accordingly.

While the original design of AMD SVM from 2005 allowed a hypervisor to run multiple guests on a single CPU without altering the guest OS, it lacked support to virtualize memory efficiently.
In 2008, AMD released a technology called ``nested-paging'' \cite{virtualization2008amd} that enables a hypervisor to virtualize memory in an efficient way.
The traditional paging hierarchy was extended with another layer, the nested layer.
Instead of just translating from virtual to physical addresses, now the translation involves two steps.
The guest pagetable, maintained by the guest operating system, translates from guest virtual to guest physical addresses,
whereas the host pagetable translates from guest physical addresses to physical addresses.
This second translation step is fully under control of the hypervisor.

\subsection{Memory Management}
\label{subsec:maintenance}
KVM leverages nested page faults as an indicator for the access load of a guest memory page.
This information is required to optimize tasks in which guest memory has to be made temporarily unavailable,
namely live migration, memory snapshots, and memory overcommitment.
Here we give a brief overview of how KVM incorporates nested page fault information in those tasks.

For live migration and memory snapshots, memory is transferred incrementally.
First, the current guest memory content is copied without stopping the guest execution.
This memory snapshot is extended in later increments by memory content that has been modified during the initial transfer.
Only if the estimated remaining transfer time falls below a predefined threshold, the guest is stopped to transfer the remaining pages.
To identify modified pages since the last transfer, QEMU instructs KVM to record a list of all pages that have been written to by the guest.
To that end, KVM removes the write access permission from all guest memory pages
and restores it only after registering the preceding write access fault.

Further, guest memory is subject to Linux standard memory maintenance operations.
Based on process page faults, memory is allocated lazily and can be swapped out to disk if unused,
thereby enabling memory overcommitment.
To do so the nested pagetable entries for the respective pages are removed,
therefore causing a nested pagetable violation if the guest tries to access them.
KVM handles the nested page fault, restores the page from the swap file and
recreates the nested pagetable entries along with the pagetable entries connecting to the QEMU userland process.

A similar method is used to lazily allocate memory for the guest upon initial startup.
To associate a memory region with the guest QEMU allocates a chunk of memory and informs KVM of the virtual memory area.
Until a page is accessed no host- or nested pagetable entries are created
other than those required for the guest kernel image and initial runtime.
If the guest accesses any additional page a nested page fault will be triggered and
handled by KVM as described in the previous paragraph.

\subsection{Virtual devices}
\label{subsec:virt_dev}
While the hardware virtualization extensions provide CPU and memory virtualization, handling device virtualization is the obligation of the hypervisor.

On x86, devices are accessed by either IO ports, memory mapped registers or by a combination of both.
Accessing IO port based devices requires the use of special instructions (e.g. \texttt{IN} or \texttt{OUT}) whereas memory-mapped devices can be accessed using normal instructions (e.g. \texttt{mov}).
If the device itself requires the CPU to handle an event, it raises an interrupt which diverts the control flow of the CPU to a specific interrupt handling routine.
To improve the overall performance, data can also be transferred without the involvement of the CPU.
The device reads or writes directly to or from main memory, allowing the CPU to perform other tasks in parallel.
The technology is commonly referred to as DMA (Direct Memory Access).
To protect against unauthorized accesses from DMA capable devices, an IOMMU can confine devices to only access configured memory regions.
Three common approaches to handling devices in a virtualized environment are passthrough, emulation, and para-virtualization.

\paragraph{Passthrough}
Using this method, one VM has exclusive access to a hardware device.
If the device provides only a memory-mapped interface, the corresponding memory pages are mapped into the guest address space via the nested pagetable.
In the case of IO ports, the \texttt{vmcb} allows configuring which IO ports are accessible directly by a guest.
If a commodity device without special virtualization extensions is passed through, only a single guest can use this device.
Some devices can also be configured to provide ,,virtual functions'', through which the same device can be used by multiple guests.
An IOMMU is usually required to contain DMA access within configured memory regions.

\paragraph{Emulation}
The hypervisor can present a virtual device to the guest.
It sets up the nested pagetable with a hole in the address space where the guest expects the memory-mapped device.
When the guest now accesses these memory ranges to interact with the device, this will trap into the hypervisor.
To perform memory access on behalf of the guest the hypervisor must know the value that should be written.
The \texttt{vmcb} will contain the fault address, i.e. the location where the data should be written, but not the value itself.
The value is usually stored in a general purpose register\footnote{There are instructions like \texttt{rep ins movs} that take the target and source address as pointers in registers, but those are not commonly used
when accessing memory-mapped devices.}.
The hypervisor must parse the instruction that caused the fault to identify the register holding the value.
As the instruction pointer locating this instruction holds a guest-virtual address, the hypervisor must first traverse the guest pagetable to get the guest-physical address of the instruction before it can parse the instruction.
As traversing the pagetable imposes a severe bottleneck for device emulation, AMD added decode assists that provide
the register location of the value in case of a nested page fault.

\paragraph{Para-virtualization}
The performance for accessing virtual devices can be enhanced using para-virtualization.
Here the hypervisor does not emulate an existing device but provides an interface of an artificial device to the guest that has no corresponding hardware device.
This has the advantage that the hypervisor and the guest can agree on an interface that encompasses the peculiarities of the hypervisor and guest communication.
For example instead of trapping writes to certain memory areas, the guest can use special instructions that cause a trap into the hypervisor.
This mechanism is called a hypercall.
In contrast to memory accesses, these hypercalls do not cause a pagetable walk by the memory management unit.
This can increase the performance of these virtual devices but requires drivers to be adapted to the hypercall interface.
\\

    Device emulation is crucial for providing basic VM functionality,
    like network connectivity, for the guest owner.
    Besides device emulation, features that are the sole responsibility of the cloud providers, such as host memory management and migration, are
    mandatory in a cloud environment.
Given that substantial modification of the deployed VM's code base goes against customer interest
    and device passthrough does not scale to larger cloud infrastructures,
    this draws a lower bound on the limitations imposed on hypervisor control over guest VMs, i.e., host memory management demands that the hypervisor has control over the second level page translations.


\subsection{Linux KVM}

\begin{figure}
    \includegraphics[width=0.8\linewidth]{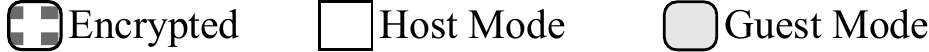}\\
    \subfloat[KVM: startup\label{fig:KVM_startup}]{
        \includegraphics[width=0.477\linewidth]{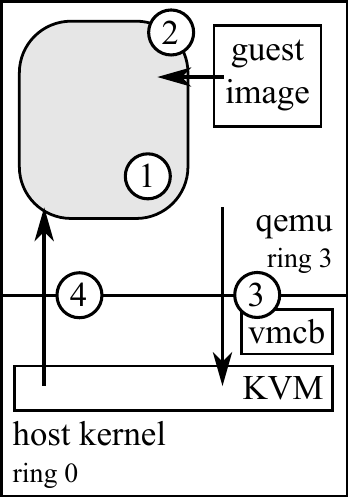}
    }
    \hfill
    \subfloat[KVM: runtime\label{fig:KVM_runtime}]{
        \includegraphics[width=0.477\linewidth]{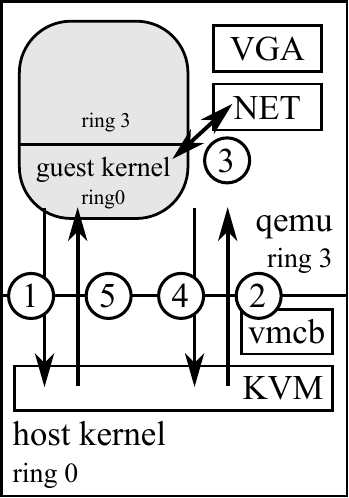}
    }
    \caption{QEMU/KVM architecture}
    \label{fig:kvm_arch}
\end{figure}

\begin{figure}
    \includegraphics[width=0.8\linewidth]{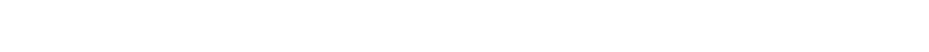}\\
    \subfloat[KVM/SEV: startup\label{fig:sev_startup}]{
        \includegraphics[width=0.477\linewidth]{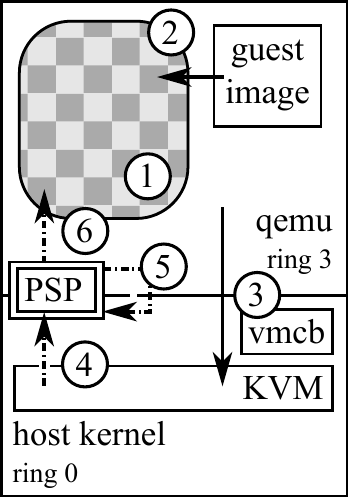}
    }
    \hfill
    \subfloat[KVM/SEV: runtime\label{fig:sev_runtime}]{
        \includegraphics[width=0.477\linewidth]{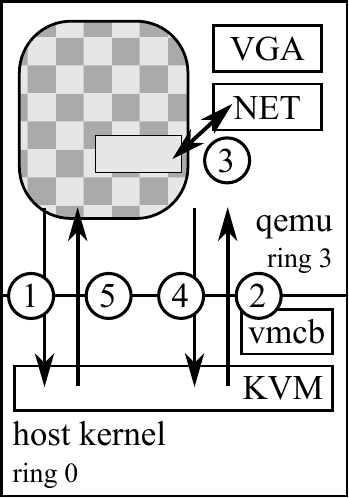}
    }
    \caption{SEV-enabled QEMU/KVM architecture}
    \label{fig:sev_arch}
\end{figure}

In the previous paragraph, we explained AMD's virtualization extensions.
We now lay out how this technology is used by the KVM hypervisor which is integrated with the Linux kernel~\cite{kivity2007KVM}.

Virtualizing CPU and memory is not sufficient because guest operating systems also need devices such as video output, network or block devices.
As a guest should not directly interfere with the hardware devices itself, they must be either multiplexed or emulated (see Section~\ref{subsec:virt_dev}).
While the KVM hypervisor is responsible for controlling the execution of guest VMs, QEMU is leveraged to handle the device virtualization.
Figure~\ref{fig:KVM_startup} depicts the initial startup of a guest VM in a KVM/QEMU setup.

First, QEMU reserves memory for the VM (Figure~\ref{fig:KVM_startup}~\circled{1}).
Then it copies the guest binaries into this reserved memory (Figure~\ref{fig:KVM_startup}~\circled{2}).
By using the \texttt{/dev/kvm} device node, the KVM module of the Linux kernel is instructed to start a new VM (Figure~\ref{fig:KVM_startup}~\circled{3}).
KVM then sets up a \texttt{vmcb} data structure incorporating the information from QEMU and issues the \texttt{vmrun} instruction to start the VM (Figure~\ref{fig:KVM_startup}~\circled{4}).
The processor now enters the guest mode, depicted in grey, and starts execution at the entry point defined in the \texttt{vmcb}.

The runtime behaviour is shown in Figure~\ref{fig:KVM_runtime}.
Upon any event that was configured in the \texttt{vmcb} to cause a \texttt{vmexit}, the CPU leaves guest mode and enters host mode again with a specific error code set in the \texttt{vmcb} (Figure~\ref{fig:KVM_runtime}~\circled{1}).
The KVM module can then either handle the exit itself or, in the case of, e.g., a memory-mapped IO operation to an emulated device, can return to QEMU which then handles the request (Figure~\ref{fig:KVM_runtime}~\circled{2}).
The emulated device can access guest memory directly to mimic DMA memory transfers (Figure~\ref{fig:KVM_runtime}~\circled{3}).
After the request was served, QEMU calls KVM again (Figure~\ref{fig:KVM_runtime}~\circled{4}), which resumes execution of the VM in guest mode (Figure~\ref{fig:KVM_runtime}~\circled{5}).

\subsection{AMD SEV}
\label{sec:sev_overview}
As indicated in Figure~\ref{fig:kvm_arch},
the hypervisor has full access to guest memory while the CPU is in host mode.
This demands that a cloud customer must trust not only the employees of the cloud provider but also the integrity of the hypervisor.
Bugs such as ~\cite{cveVirtualBox, cveXen, cveVmware, cveHyperV, cveVenom} can be used by a malicious tenant to attack the hypervisor itself and thereby gain access to assets of other tenants residing on the same physical machine.

SEV protects guest memory via encryption.
The guest specific memory encryption key will never be exposed to the hypervisor.
It is only accessed by
a secure coprocessor and the memory controller that handles the encryption and decryption transparently.
The coprocessor which was added to SEV-enabled CPUs (the ,,Platform Security Processor''~\cite{sme2016},
indicated as \textit{PSP} in Figure~\ref{fig:sev_startup}),
handles key management and is responsible for configuring the correct guest key within the memory controller.

Figure~\ref{fig:sev_arch} shows how the classical KVM architecture looks on an SEV-enabled system.
Like detailed in the previous Section,
QEMU communicates with the KVM module to  prepare the VM for launch
(Figure~\ref{fig:sev_startup} \circled{1} to \circled{3}).
To enable SEV for the newly allocated VM it's memory must first be encrypted.
The host kernel calls the coprocessor to initiate the encryption of the
VM memory using a threefold command sequence, \texttt{LAUNCH\_START}, \texttt{LAUNCH\_UPDATE} and \texttt{LAUNCH\_FINISH}
(Figure~\ref{fig:sev_startup}~\circled{4}, \circled{5} and \circled{6}).
By using this command sequence, the hypervisor ensures that the firmware generates an encryption key
unique to the VM (\texttt{LAUNCH\_START}),
encrypts the memory and records a launch receipt of the VM used for remote attestation (\texttt{LAUNCH\_UPDATE} and \texttt{LAUNCH\_FINISH}).
After the encryption of guest memory is completed the firmware provides the recipe to the hypervisor
to be passed on to the customer.
This recipe includes measurements of the guest image and platform authentication data, which
allow the customer to verify that the VM memory was encrypted and initialized correctly.
If a customer judges the recipe or the contained measurements to be faulty,
he can choose to withhold the provisioning of secrets to the VM.

Each VM uses its unique cryptographic key that is loaded by the secure processor when the corresponding VM is scheduled.
Once a guest enables paging, it can mark individual data pages as either \textit{shared} or \textit{private} by setting a physical address bit (the enCrypted- or C-bit) in its pagetable.
Memory pages marked as \textit{private} are encrypted using AES with the guest specific key and pages marked as \textit{shared} are either not encrypted or encrypted with the hypervisor key and can thus be used to exchange data with the hypervisor.
The C-bit of the guest pagetables has precedence over the C-bit of the hypervisor controlled second level pagetables
to secure the page protection configured by the guest VM.
    In addition to the memory protection mechanism, AMD offers tenants the ability to enforce guest policies.
    Policy configuration includes amongst others
    the option to disable debug capabilities of the hypervisor towards the guest VM.

Figure~\ref{fig:sev_runtime} shows the system configuration during runtime.
The secure coprocessor (\textit{PSP}) is not shown, as it is used mainly during VM startup.
The steps composing the runtime behaviour under SEV (Figure~\ref{fig:sev_runtime}
\circled{1}, \circled{2}, \circled{4} and \circled{5})
do not differ from the non-SEV configuration.
This is due to the fact, that cryptographic operations are handled transparently by the memory controller,
while the secure processor handles key management without the involvement of the hypervisor or the VM.
However to facilitate DMA memory transfers (Figure~\ref{fig:sev_runtime}~\circled{3})
similar to a classical setup,
the guest is tasked to configure shared memory regions, which are exempt from encryption.

\subsection{SEV - Encrypted State}
SEV - Encrypted State (SEV-ES) is an extension to SEV that additionally encrypts the guest state,
including the general purpose register, using the guest specific encryption key.
When the CPU leaves the guest mode, all general purpose registers, as well as the guest saved state, are encrypted.
A \texttt{vmexit} event is now classified as either an ``Automatic Exit'' (AE)  or a ``Non-Automatic-Exit'' (NAE)
depending on the exit reason.
Any asynchronous event, e.g. an interrupt, is classified as an AE.
AE events do not require the hypervisor to read the guest state, which can therefore be encrypted by the secure processor.
AES events, on the other hand, are events that potentially require the hypervisor to read the guest state.
If such an event occurs, the control is not transferred to the hypervisor.
Instead, a new exception is raised in the guest, the ``VMM Communication Exception'' (\#VC).
The exception handler of these exceptions in the guest can now decide whether to provide the hypervisor with access to the guest state, or not.
To exchange data a shared memory region is used, called ``Guest-Hypervisor Communication Block'' (GHCB).
Data to be shared with the hypervisor must be copied to that memory region.
After copying the values to the GHCB, the guest executes \texttt{vmgexit} to transfer control to the hypervisor.
Nested page faults are usually AE events, i.e., the hypervisor does not need to read guest state to handle these events.
Therefore there is no \#VC exception triggered in these cases.
In order to allow the hypervisor to provide emulated devices for the guest, the hypervisor can enforce certain nested page faults to be NAE events.
To do so, the hypervisor sets a reserved bit in the nested pagetable.
Faults caused by accesses to these pages are treated as NAE events, and instead of transferring control to the hypervisor, a VC exception in the guest is raised.


\section{AMD SEV Security Considerations}
\label{sec:sev_considerations}
While guest memory is protected from direct hypervisor access by encryption, other security-critical components are not protected at all.
By examining the AMD SEV documentation~\cite{sev2016,sme2016} 
and publicly available comments from AMD employees~\cite{lkmlGPREGS}, we found that for a system without SEV-ES:

\begin{enumerate}
    \item The general purpose registers are not encrypted upon a \texttt{vmexit}~\cite{lkmlGPREGS}.
    \item The \texttt{vmcb} is subject to manipulation by the hypervisor~\cite{lkmlGPREGS}.
    \item{The hypervisor can access encrypted guest memory due to the lack of memory authentication~\cite{sev2016,sme2016}.}
\end{enumerate}

Under a system, which also implements the SEV-ES extension only the third point remains valid.

\paragraph{General Purpose Registers}
Whenever the CPU switches from guest mode to host mode, the general purpose registers of the guest are exposed to the hypervisor.
As the guest itself cannot control 
when the CPU transfers to host mode, these registers can contain potentially confidential data. 
If such an exit occurs e.g. while the guest is generating an RSA key pair, 
the key components might be exposed to the hypervisor.

\paragraph{VMCB}
As mentioned in Section~\ref{subsec:x86_virt}, the \texttt{vmcb} is used 
to control the execution and state of the guest.
The \texttt{vmcb} is therefore crucial to guest integrity and exposes the content of privileged guest registers.
Among these registers is the instruction pointer of the guest 
which allows the hypervisor to govern guest control flow.

\paragraph{Memory Authentication}
The memory is encrypted, but it is otherwise not protected from access.
This enables the hypervisor to inject faults into the guest or to  capture and replay private guest memory.\\

Later sections will lay out how these design issues can be leveraged by a malicious hypervisor to
a) gain shell access to a guest, 
b) read protected guest memory and
c) fully revert any memory protection configured by the tenant.
While SEV-ES successfully protects the general purpose registers and the \texttt{vmcb}, the guest memory can still be accessed by the hypervisor, though the hypervisor can only access encrypted pages.
This mitigates attack vectors b) and c).
Still, as we will show in later sections there is no easy way to prevent a) without sacrificing guest transparency or
impacting classic cloud functionality, such as migration and the memory management features of the hypervisor.

\section{Attack Model}
\label{sec:attack_model}
In this section, we describe our attack model, which is based on the AMD SEV security properties (detailed in Section \ref{sec:sev_considerations}).

We assume that a customer successfully deployed his VM on an AMD SEV-enabled system.
During startup, we also assume that the hypervisor is uncompromised and compliant with the AMD SEV 
specification~\cite{sev2016}.
This means the customer was able to attest the correct setup of his VM using the receipt provided by the hypervisor.
From this point on the VM is protected by AMD SEV.
Neither the hypervisor nor someone with physical access to the cloud infrastructure is able to read the designated private memory regions of the protected guest.

Then, during runtime, an attacker was able to compromise the hypervisor, thereby gaining root access 
to the host system.
The described scenario is likely, as incidents of the past show~\cite{cveVmware, cveHyperV, cveXen, cveVenom}.
We also assume that the attacker has knowledge of the target system with regards to the versions of the kernel and userland processes.
As the cloud provider itself often provides the guest images, this is also likely.
We assume that the encryption scheme in use produces the same encrypted data if the input, 
key and host physical address are identical, 
similar to other symmetric linear memory encryption schemes.
Further, we require, that no integrity check is performed on encrypted data,
In addition to that, we initially assume access to nested pagetables, \textit{vmcb} and guest register state,
which we later restrict only to nested pagetable access
for the replay attack.

\section{Attacks against Encrypted Virtualization}
\label{sec:attacks}
We now present three attacks against VMs under a compromised hypervisor.

The first two attacks presented in Section~\ref{sec:control_struct} are directed against the proposed design of SEV without the optional feature SEV-ES, 
which allows the hypervisor to extract and control guest state through 
the unencrypted guest control block and registers.
Amongst other security concerns for the tenant, this flaw can be used
to decrypt guest memory including the internal address mapping, 
as we will show in our first attack in Section~\ref{sec:decryption}.
Building upon this capacity we describe in Section~\ref{sec:disable_prot} how the memory protection 
configured by the tenant can be deactivated, without notifying the guest.
After deactivation of memory protection further exploitation,
like arbitrary code execution is trivial to execute as the hypervisor now has full access to guest memory.

The third attack already takes the presence of SEV-ES into account, 
which protects the integrity of the guest control block and registers in the face of a malicious hypervisor.
We however show in Section~\ref{sec:replay_attack_design} 
that protecting those structures alone is not sufficient.
If the hypervisor is in control over guest memory allocations through nested paging,
it can use this capability to launch a replay attack.
We prove this claim by launching an attack against an OpenSSH server running in the protected guest VM to gain 
access at potentially high privilege levels.
\subsection{Attacks based on exposed Guest State}
\label{sec:control_struct}

In this section, we present two attacks against an encrypted guest,
facilitating hypervisor access to the guest control block and registers.
First, we describe a method to exploit guest control flow
to read and write arbitrary memory areas of a running guest in decrypted form.
Based on this primitive,
we construct an advanced attack to
disable guest memory protection as documented in~\cite{sev2016} altogether.

\subsubsection{Accessing Protected Memory}
\label{sec:decryption}
Given a system which is capable of encrypting guest memory as described in ~\cite{sev2016},
we now describe how a malicious hypervisor can coerce an encrypted guest into leaking arbitrary memory content.
The methods for reading and writing protected guest memory are symmetric,
therefore we restrict this section to the description of the memory read primitive.

During guest execution, the memory of the active VM is transparently decrypted by the memory controller.
Memory content which, in this state, is transferred
into unencrypted areas like the \texttt{vmcb}, registers or shared memory,
will be exposed to the hypervisor whenever guest execution is interrupted.
Our attack induces an interruption of the guest execution,
right after protected data has been transferred from an attacker controlled memory location into an unencrypted register.
To divert guest control flow, we set the guest instruction pointer before guest re-entry
to the guest virtual address of a suitable instruction sequence.
Shortly after the read instruction we force a \texttt{vmexit} to read the decrypted data
from the register.

\begin{figure}
    \centering
\begin{lstlisting}[caption=Read Instruction Sequence, label=decrypt_gadget]
mov edi, dword ptr [rbx]
hlt
\end{lstlisting}
\end{figure}

The instruction sequence is required to
end with a trappable instruction and to contain an indirect memory read.
Listing~\ref{decrypt_gadget} shows the sequence of instructions, which we used to launch the attack.
We extracted this sequence statically from the guest kernel binary,
for which we used a modified tool for ropchain generation, called ROPGadget \cite{ropgadget}.
The code snippet reads four bytes from guest memory into the register \texttt{eDI},
before a \texttt{vmexit} is induced by the instruction \texttt{hlt}.
The malicious hypervisor can then conveniently take the decrypted word from the general purpose register.
Listing~\ref{clts_exit} shows the respective exit handler, which
the hypervisor could use to handle this particular \texttt{hlt} trap condition.
To decrypt an arbitrary section of guest memory,
the exit handler re-sets the guest instruction pointer to the guest virtual address of the instruction sequence and
the guest register \texttt{rBX} to the guest virtual address of the protected memory to be read.

The diversion of guest control flow can be initiated at any point during host execution.
To resume normal guest execution after the attack,
the guest registers which are clobbered by the decryption
are saved in the host environment and later restored after the final memory element has been read.


\begin{figure}
    \centering
\begin{lstlisting}[caption=HLT Exit Handler, label=clts_exit]
int new_handle_hlt(struct vcpu* vcpu) {
    u64 rip, edi;
    rip = rip_read(vcpu);
    if(rip == DECRYPT_HLT_INS && decrypting) {
        edi = register_read(vcpu, VCPU_EDI);
        // process decrypted data in EDI ...
        register_write(vcpu, VCPU_RBX, current_addr);
        current_addr += 0x4;
        if(current_addr < last_addr) 
            rip_write(DECRYPT_HLT_INS);
        return 0; 
    } else {
		// handle normal hlt exit ...
    }
}  
\end{lstlisting}
\end{figure}

\paragraph{Locating the Instruction Sequence}
The recent introduction of kernel address space layout randomization (\textit{KASLR}) complicates our attack.
Now the instruction sequence cannot simply be obtained from the guest kernel binary.
Instead, we only obtain the offset of the sequence within the kernel text section via static analysis.
The offset is then added to the dynamic load address of the kernel text section,
which is randomly initialized during the boot process of the VM.
To compute the load address of the kernel text section,
we compare control registers exposed through the guest control block,
pointing to kernel functions inside the guest's virtual address space.
Specifically, we subtract the virtual address of the system call entry function \texttt{entry\_SYSCALL\_64} of the running guest
from the system call entry address of the non randomized kernel image.

\subsubsection{Disabling Memory Protection}
\label{sec:disable_prot}

In this section, we describe how encryption can be disabled
for individual guest memory pages or even for the complete guest memory space.
The attack is based on manipulation of guest internal pagetable entries.

First, we will describe how we access those entries, even though they are assumed to be located in private guest memory and thereby to be encrypted.
To access pagetable entries within the protected guest, we first 
read the physical pagetable address of the currently active process from the \texttt{cr3} register value
stored in the \texttt{vmcb}.
We then use the method described in the previous section to read pagetable entries from protected guest memory.
As the read primitive can only operate on guest virtual addresses,
we access the pagetable data via the direct physical memory map (referred to as \textit{physmap}).
The \textit{physmap} is a contiguous mapping of the physical RAM into the virtual address space of the kernel.
The virtual base address of the \textit{physmap} is stored in the kernel variable \texttt{page\_offset\_base}
which is located at a constant offset from the dynamic load address of the kernel's text section.
We use the read primitive with the adjusted offset of the \texttt{page\_offset\_base} variable to read its value from guest memory.
To access the pagetable entries, we add the physical pagetable base address to the virtual base address of the \textit{physmap}.
Using the write primitive, we are now able to overwrite and add pagetable entries arbitrarily,
using the adjusted guest virtual address of the pagetable base as the target location.

Even though we are now able to change the pagetable bit controlling page encryption,
clearing the C-bit from a guest pagetable entry will only disable the transparent de-or encryption
on subsequent memory read or write accesses.
Thus the attacker is required to allocate new unprotected (shared) pages of memory
and to copy the protected (private) data into the newly allocated areas.

\begin{figure}
    \centering
    \subfloat[Duplication\label{fig:pt_moda}]{
        \includegraphics[width=0.9\linewidth]{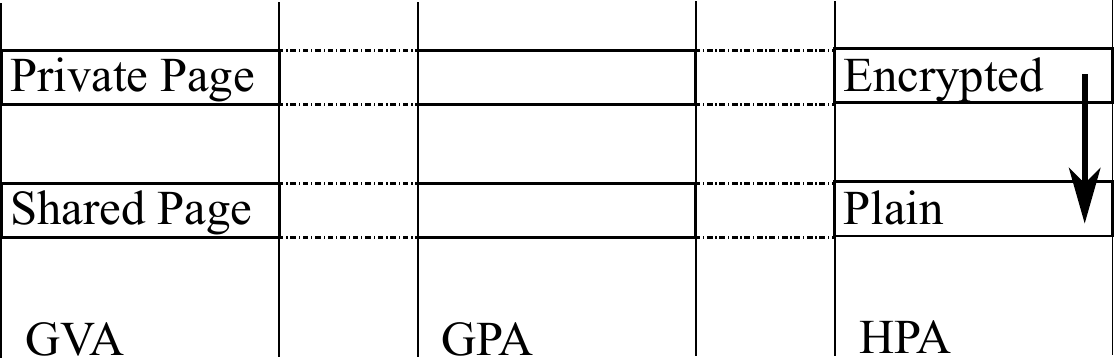}
    }
    \hfill
    \subfloat[Replacement\label{fig:pt_modb}]{
        \includegraphics[width=0.9\linewidth]{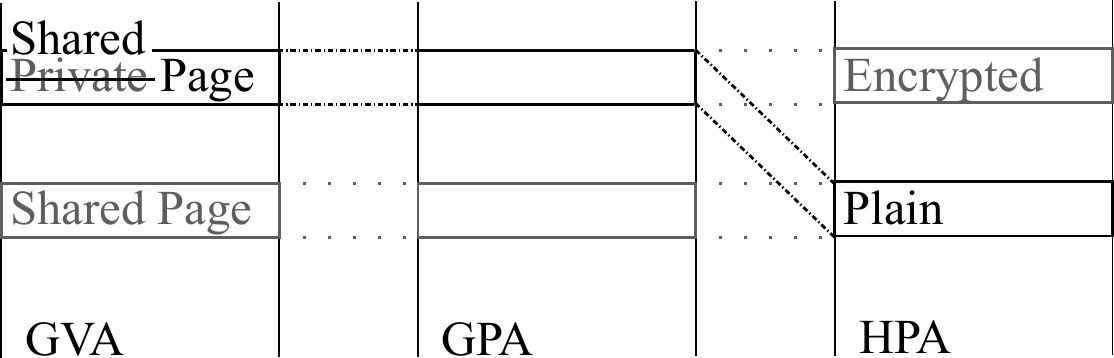}
    }
    \caption{Pagetable Modification}
    \label{fig:pt_mod}
\end{figure}

Figure~\ref{fig:pt_mod} gives an overview of how an attacker can deactivate
the protection of guest pages under an arbitrary guest pagetable entry,
without notifying the guest.
The process can be split up in two phases, first, duplication and then, replacement.
In the duplication phase, the protected data is transferred into newly allocated memory
as seen in Figure~\ref{fig:pt_moda}.
During the replacement phase, the guest pagetable entry
is modified to deactivate the protection,
while the nested pagetable entry is redirected to the new data, as shown in Figure~\ref{fig:pt_modb}.

Now to actually decrypt an amount of guest pages, the according amount of pages has to be reserved in host memory.
Then using the read primitive, the guest pagetables are browsed for an unallocated slot matching the original entry level.
Similarly, an empty slot in the second level pagetables is located.
New pagetable entries are created from the host physical, guest physical and guest virtual addresses.
The write primitive has to be used to add the entry connecting the guest virtual to the guest physical address
in the guest pagetables.
Using the read primitive again, the protected guest memory is read and directly written
into the newly allocated area by the host.
Finally, the pagetable entry of the original protected mapping is modified to clear the C-bit,
while the nested pagetable entry is redirected to point to the newly written data.

\subsection{Attack based on Nested Pagetable Control}
\label{sec:replay_attack_design}

Systems that implement the optional SEV-ES feature will be protected against the previously mentioned attacks.
Based on these revised security properties we construct a third attack,
which relies only on control over nested pagetable structures and interrupt injection capabilities of the hypervisor.
We leave the detailed discussion about the necessity of the
latter two capabilities for guest transparent VM encryption in a cloud environment to Section~\ref{sec:discussion}.

SEV-ES limits the hypervisor's control over a guest.
Namely, the feature restricts access to the VM control block and
forces the encryption of guest general purpose registers upon a \texttt{vmexit}.
Here we explain the protection achieved through deploying these mitigations to motivate the next attack.

Limiting access to the VM control block prevents the execution of the previous attacks on several levels.
The leakage of kernel function pointers is prevented; therefore the guest internal address mapping is not revealed.
Whether an instruction like \texttt{hlt} traps into host mode is controlled via a bitmap contained in the \texttt{vmcb}.
Therefore the number of instruction sequences suitable for misuse as read and write primitives can be limited by
controlling the configuration of this bitmap.
Further, the capability of the hypervisor to manipulate guest control flow is restricted,
as the instruction pointer, which is also part of the \texttt{vmcb}, can be protected from malicious modification.
The encryption of guest control registers will handicap the application of read and write primitives by
impairing the control over the address of accessed memory as well as the exposure of the decrypted data.
We argue that limiting  hypervisor control over physical memory assignment would prevent memory overcommitment as well as
any dynamic load balancing or migration efforts.
In fact, we assume this capability to be crucial in a cloud environment.
Comparably critical is the ability to inject virtual interrupts for device virtualization.

We now describe how a malicious hypervisor
can launch a replay attack against a VM running in a protected environment,
which uses the optional SEV-ES~\cite{amd64architecture} feature
in addition to the protection mechanisms proposed by SEV~\cite{sev2016}.

First, we give a brief overview of replay attacks and explain
how we can attack an OpenSSH server running in an unprotected guest by replaying login credentials.
Next, we describe how we can infer the correct location and time to capture and replay guest memory
\textit{without} insight into the guest memory content,
by observing memory access and system call patterns of the guest.
Finally, we describe the steps necessary to implement the attack against an encrypted VM.
We conclude with an evaluation of the presented attack.

\subsubsection{Replay Attacks}
\label{sec:replay_attacks}
On a high level, replay attacks exploit the lack of data versioning and authentication,
which allows an attacker
(in our case a malicious hypervisor)
to eavesdrop on the exchange of valid authentication tokens and replay them to pose as the original communication partner.
For OpenSSH, we identified the function \texttt{userauth\_passwd}, shown abbreviated in Listing~\ref{fig:openssh},
as a suitable target.
In \autoref{lst:packet_get_string} a password string is read
from the network buffer via \texttt{packet\_get\_string}
and stored on the stack.
The password string is then validated at \autoref{lst:auth_passwd} by \texttt{auth\_passwd}.
After validation, the \texttt{password} is removed from memory at \autoref{lst:memset}.

To launch the attack against the OpenSSH server executing in a \textit{unencrypted} guest,
the hypervisor captures the guest page containing the credential data in between
lines \ref{lst:packet_get_string} and \ref{lst:auth_passwd}.
The attacker then initiates a new connection.
After the server receives credentials from the attacker controlled client,
the hypervisor replaces the invalid credentials of the attacker, with the data captured in the previous step.
The validation of the restored password will then succeed and thereby grant access to the attacker controlled client
at the privilege level of the connecting user.

\begin{lstlisting}[breaklines,label=fig:openssh,caption=userauth\_passwd,numbers=left,numbersep=1pt,escapeinside={@}{@}]
static int userauth_passwd(Authctxt* authctxt) {
       char *password, *newpass;
       // ...
       change = packet_get_char();
       @\label{lst:packet_get_string}@password = packet_get_string(&len);
       // ...
       if (change)
              logit("password change not supported");
       @\label{lst:auth_passwd}@else if (PRIVSEP(auth_password(authctxt, password)) == 1)
              authenticated = 1;
       @\label{lst:memset}@memset(password, 0, len);
       xfree(password);
       return authenticated;
}
\end{lstlisting}

\subsubsection{Inferring Memory Content}
\label{sec:inferring}

In a classical replay scenario, the hypervisor can monitor memory content
to identify location and state of the memory region to be captured and later replayed.
If the guest memory is encrypted, the main challenge is to infer those parameters indirectly.

In this section, we describe how we identify when and where to capture and later replay a memory page
without insight into its content.
The key intuition behind our approach is that memory content can be inferred through the
access patterns to individual pages, which we express through system call sequences.
First, we explain how we extract information about system calls issued by the guest.
Next, we describe how the sequence of system calls issued by the guest is combined
with the sequence of writes to guest memory
to identify the location of selected data structures as well as their state.

\paragraph{Trapping System Calls into the Hypervisor}
To record a sequence of system calls issued by the guest
we need a mechanism to trap into the hypervisor when a guest user process tries to execute a system call.
It is important to note that we can extract system call information without access to the guest register or control state.
Instead, we remove the execute permissions on the guest memory page
containing the system call entry function \texttt{entry\_SYSCALL\_64}
as well as the pages containing the system call handler routines.
Thereby we enforce an exit to the hypervisor whenever system call execution is initiated.
By examining the fault address, the hypervisor can determine which handler caused the fault.
Initially, only the system call entry page is protected;
if a \texttt{vmexit} is induced by guest execution of this page, we restrict access to the handler pages and
restore execute permissions on the entry page.
Similarly, if a \texttt{vmexit} is induced by guest execution of one of the handler pages
we restore execute permissions to all handler pages while restricting access to the entry page.
This procedure is necessary, to enable the re-execution of the faulting instruction in the guest.

To remove execute permission from guest pages containing the respective functions,
we first need to locate them in guest physical memory.
Due to \textit{KASLR}, the physical load offset of the kernel text section is randomly initialized during the VM boot process.
We therefore employ a similar method as described in Section~\ref{sec:decryption}
to adjust the guest physical addresses of the system call entry and handler functions accordingly.
To obtain a point of reference from which to compute the physical load offset of the kernel text section,
the hypervisor can trigger the immediate execution of known functions, like interrupt handler routines, by the guest.
By previously marking \textit{all} guest memory pages as non executable through the nested pagetables,
the guest will immediately fault, revealing the physical address of the triggered function
through the fault metadata provided to the hypervisor.
The random physical load offset of the kernel text section is then calculated by subtracting the fault address
from the physical address of the function, obtained from a non randomized kernel image.

\paragraph{Combining System Call and Write Sequences}
\begin{figure}
    \centering
    \includegraphics[width=.8\linewidth]{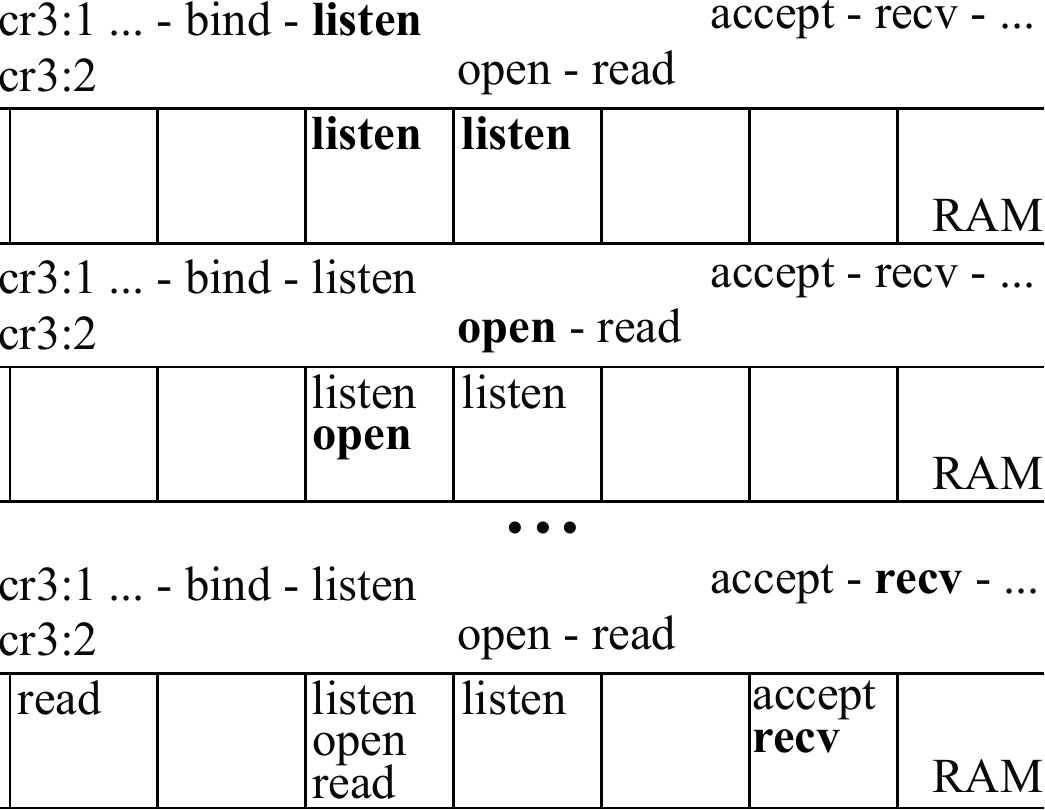}
    \caption{System Call Sequences on Guest Memory}
    \label{fig:tracing}
\end{figure}
Based on the recorded system call sequence,
the hypervisor can reason about the state of guest execution.
However, we still lack the ability to identify which of the many guest memory pages that are written
continuously, contains the data selected for replay.
To that end, we cross-reference the sequence of guest memory writes with the system call information
by storing a sequence of system calls for each page
that preceded a write access to the respective page.

We now explain our approach via a simplified example.
Figure~\ref{fig:tracing} shows an excerpt of guest execution with two concurrent processes.
The processes are identified by their according \textit{cr3} values (either 1 or 2).
Each process performs a number of different system calls, whereas the most recent one is highlighted in bold font.
Guest pages are subsequently marked with the system call identifier that was last recorded
before a write access occurred.
To record those we intercept the guest on memory write access
in addition to system call execution.
Upon a \texttt{vmexit} induced by execution of a system call handler,
we now also remove write permission from all guest pages.
Each subsequent write will now trap to the hypervisor,
where we firstly restore write permissions for the respective page
to allow for the re-execution of the faulting instruction
and secondly, mark the accessed page with the last recorded system call identifier.
On each memory write we then evaluate these sequences for all guest pages
to infer whether a specific page currently contains the data selected for replay.


\subsubsection{Replay Attacks against Encrypted VMs}
\label{sec:enc_replay_attack}

In this section we first describe the four phases composing our replay attack,
namely offline analysis, tracing, capture, and replay.
We then illustrate these steps
by describing our procedure to replay OpenSSH login credentials to gain access
to an encrypted VM
at the privilege level of the connecting user.

\paragraph{Offline Analysis}
The first stage is an offline analysis of the target application to determine possible replay attack vectors.
Currently, we do this manually and on a per-application basis.

\paragraph{Tracing}
To determine the location of the credential data structure in encrypted guest memory,
we first trace system call and memory access patterns of an unencrypted guest running an identical OS and target application.
This allows us to scan the unencrypted guest memory for the selected data continuously.
If the selected data is detected in a guest memory page, we store the respective access pattern.
Due to interrupts, scheduling and input from external sources,
execution paths and therefore system call sequences might differ slightly.
We account for this by collecting multiple traces and simply extracting the
longest trailing sequence occurring in most of the traces.
The sequence starts at the syscall after which the memory is replayed.
From there we trace backward along all collected syscall sequences until one of them diverges
while discarding sequences whose length falls under a predefined threshold.

\begin{figure}
    \centering
    \subfloat[Capture\label{fig:ept_capture}]{
        \includegraphics[width=0.477\linewidth]{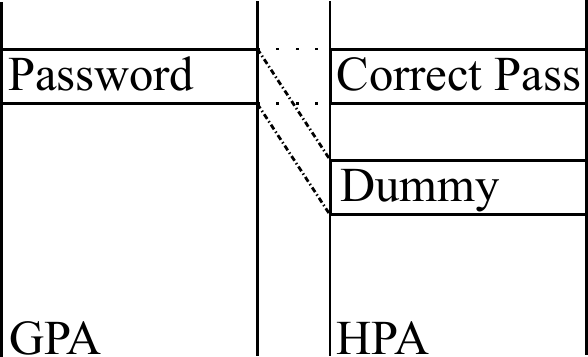}
    }
    \hfill
    \subfloat[Replay\label{fig:ept_replay}]{
        \includegraphics[width=0.477\linewidth]{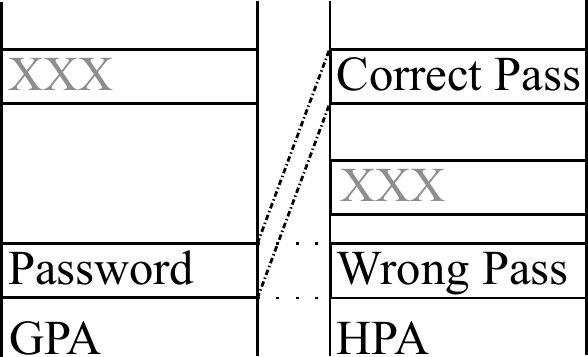}
    }
    \caption{Nested Pagetable Modification}
    \label{fig:nested_mod}
\end{figure}
\paragraph{Capture and Replay}
In the capture and replay phases, we compare the collected sequence against those generated by the encrypted guest.
If the system call sequence of a guest page matches the reference
we conclude that the encrypted page contains the selected data and
proceed to capture or replay the contained data respectively.

In accordance with the encryption scheme described in Section~\ref{sec:attack_model},
the cipher text produced by the memory encryption algorithm
is influenced not only by the content but also by the host physical address of the memory page.
Therefore the replayed data has to be placed at the same host physical address (HPA) as the captured data.
This can be achieved by manipulating the nested pagetables,
which control the mapping of the guest physical address (GPA) to its HPA as shown in Figure~\ref{fig:nested_mod}.
In Figure~\ref{fig:ept_capture} the guest memory page, containing the valid credentials
\textit{"Correct Pass"} has been identified for capture.
The nested pagetable entry connecting the GPA to the HPA of this page is then
modified to redirect the GPA to a newly allocated page (\textit{Dummy}).
This removes the captured data from the guest's address space so that it will not be overwritten.
During the replay phase described in Figure~\ref{fig:ept_replay} the
nested pagetable is again modified to redirect the GPA from the HPA containing the
\textit{"Wrong Pass"} to the HPA of the previously captured page still containing the \textit{"Correct Pass"}.
Using this method the minimum size of data that can be replayed is a single page (usually 4KB),
because the address translation can be changed only on page granularity.

\begin{figure}
    \centering
\includegraphics[width=0.9\linewidth]{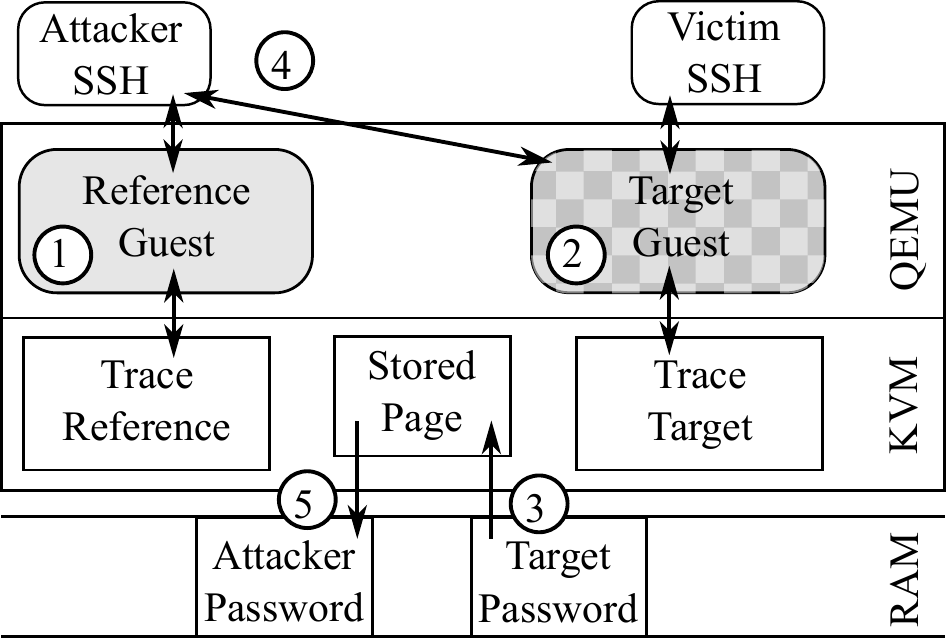}
\caption{Replay Attack Overview}
\label{fig:replay_overview}
\end{figure}
Figure~\ref{fig:replay_overview} gives an overview of our replay attack.
Here we assume that offline analysis has already identified data and state for capture and replay.
We collect a reference sequence of system calls (\textit{Trace Reference})
for the page containing the identified data,
by initiating SSH password logins \circled{1}
to an OpenSSH server running in an unencrypted guest
(\textit{Reference Guest}) with the same software configuration as the target.
Next, we wait for an incoming SSH client connection to the
protected guest (\textit{Target Guest}) \circled{2},
while continuously comparing the access patterns of the protected guest (\textit{Target Trace})
against the reference.
If an SSH client, authenticates itself to the server via password, the page containing the credential data structure (\textit{"Target Password"})
is identified and the content of the page is stored \circled{3}.
We then re-initiate a password login to the protected guest
from the attacker controlled SSH client \circled{4}.
To grant access to the attacker controlled client, the hypervisor modifies the nested paging structures
to redirect the page containing the invalid credentials of the attacker (\textit{"Attacker Password"})
to the stored page \circled{5}.

\subsubsection{Impact}
To show the effectiveness of the replay attack, we evaluate it by exploiting OpenSSH version 6.7p1-5+de running in a VM.
The test was conducted on a AMD Phenom II X4 965 processor with 4GB RAM.
As the host operating system, we used Linux with kernel version 4.4.0 and QEMU version 2.7.50 for communication with the KVM driver module.
For the evaluation, we disabled symmetric multiprocessing on the host system.
The guest was configured with 512MB RAM and ran kernel version 4.9.0-rc5 with the full range of \textit{KASLR} options enabled.
As AMD SEV is not available at the time of this writing,
we substitute an unencrypted VM as our target.
We argue that the results apply to a future real SEV setup,
because none of the data structures required for the attack will be obscured
even if SEV is enabled, according to the currently available documentation.

The effectiveness of our attack is best classified by
the number of successful logins to the target guest
that have to be observed on average, before successful execution of the described replay attack.
The success rate hinges on the accuracy of page identification via system call and memory access patterns
as well as on the structure of the page selected during offline analysis.

\paragraph{Page Structure}
We found that the offset of the credential data within a memory page varies between four separate values.
The distribution of the offset values is shown in Figure{~\ref{fig:offsets}.
To determine this distribution
we initiated 387 SSH password logins using a unique password to simplify the identification
of the page.
By examining the collected traces,
we determined that the specific location cannot be extracted
from the system call access sequence.
Further, we discovered that replaying captured data over a
page with mismatching offset will terminate the guest process handling the login.
However, termination of the process spawned by the SSH server to handle the connection will not impact the
functionality of the guest, since it is immediately respawned by the parent process.
Unsuccessful replay attempts will require the re-initiation of a new capture
because the captured page was mapped back into the guest's memory space.
Overwriting or removing a mapped guest physical page will result in unpredictable behaviour,
unless the content of the page is known.

\paragraph{Trace Accuracy}
To improve the coverage of guest execution paths and thereby the trace accuracy
we collect multiple system call sequences.
From those, we extract the sequence of system call identifiers
that identifies the greatest number of the collected traces correctly.
To measure the trace accuracy, we collected 387 traces to compute the reference sequence.
We then proceeded to initiate 2155 SSH connections to the VM and
matched the generated traces against the reference.
To verify the correct identification of the page, we choose a unique string as the password
and tested whether the guest page identified by the reference sequence contained this string
and whether the data structure of the page matched the data structure of page selected for replay.
The achieved identification accuracy for the page containing the credential data was \textbf{86\%}.
We encountered no false positives during our test, which is important
because the remapping of a falsely identified guest page during capture
will most likely have an adverse effect on the guest.

\begin{figure}
    \centering
    \includegraphics[width=.9\linewidth]{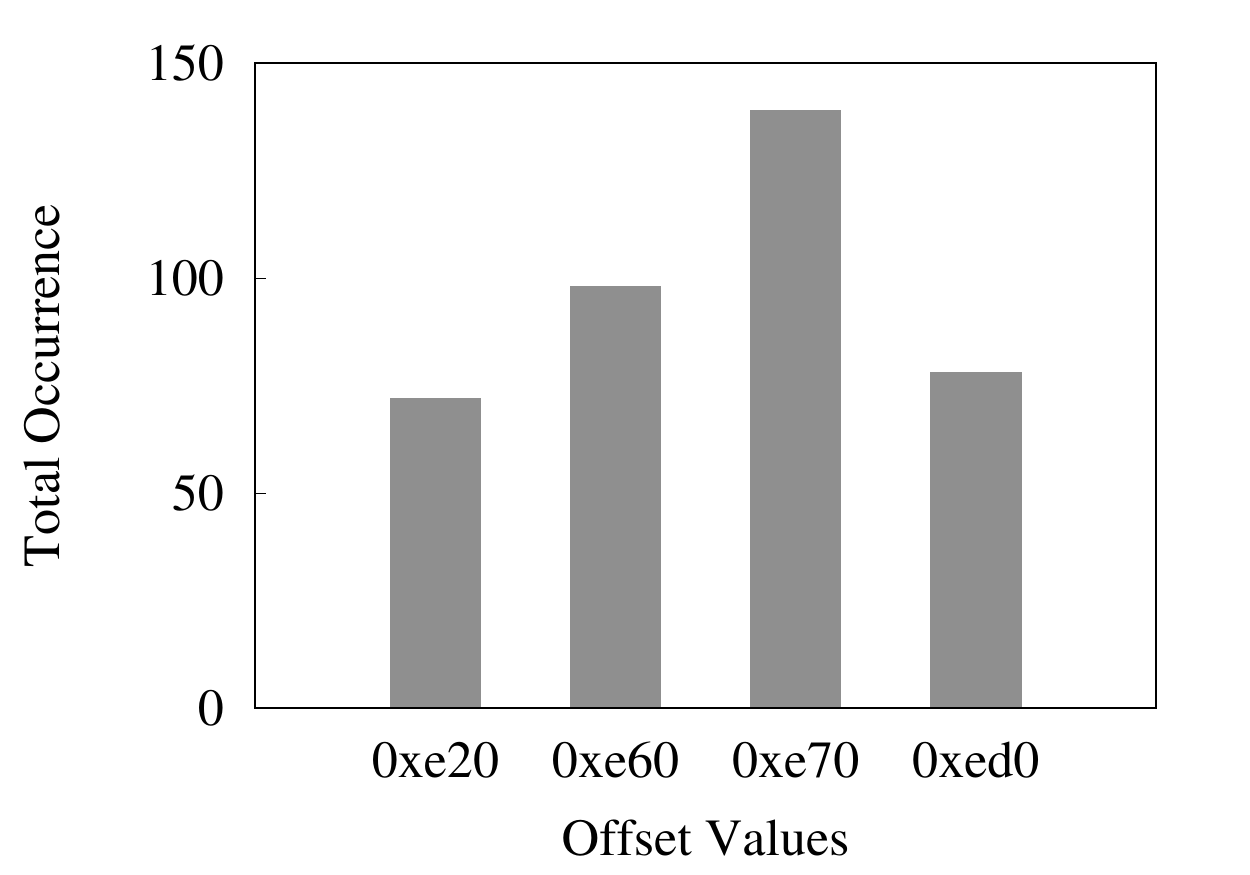}
    \caption{Credential Offsets}
    \label{fig:offsets}
\end{figure}

\paragraph{Success Rate}
To measure the overall success rate, we compare the number of observed valid logins
against the number of times, the attacker was granted access to the system.
To that end, we run two identical VMs (reference and target) and
extract a reference sequence from 387 observed logins to the reference VM.
While using this reference to identify the password data in the target VM's memory,
we initiated 2155 SSH password logins with valid login credentials to the target VM.
This resulted in 505 successful replays; therefore the success rate of the attack is \textbf{23\%}.
The result is consistent with the measurements of data location distribution and trace accuracy.
Especially the variation of credential offsets within a memory page limits the possible success rate to maximally 25\%.
However we argue that this factor can be alleviated by a more thorough investigation of the target software stack,
to identify data structures suited for replay, with less varying location offsets.


\section{Discussion}
\label{sec:discussion}
In the previous sections we laid out the details how a malicious hypervisor can exploit design issues in AMD's upcoming Secure Encrypted Virtualization technology to a) to gain access to a guest, b) to read encrypted guest memory and c) to fully revert
any memory encryption configured by the tenant.

In this chapter, we discuss possible mitigations to these threats and
evaluate their projected impact on performance and usability.


\subsection{Mitigations}
Pure software changes can not eliminate the design issues discussed in Section \ref{sec:sev_considerations}.
To thwart the attacks presented here we propose the following design changes for future versions of SEV:

\begin{itemize}
    \item{Encrypted general purpose registers}
    \item{No access to the \texttt{vmcb} after an initial configuration}
    \item{Memory protection against hypervisor access}
\end{itemize}

While the presented mitigations against the first two attacks
induce a bearable degradation of guest transparency and performance,
the replay attack remains difficult to mitigate given these demands.

\paragraph{Access to general purpose registers} 
The general purpose registers must never be visible to the hypervisor as they leak sensitive guest data on any \texttt{vmexit}.
A guest does not have control over exits to the hypervisor,
thus the encryption of general purpose registers must be enforced by the hardware.
The encryption imposes another difficulty as certain guest operations require the hypervisor to read the general purpose registers.
For example, when the guest writes data to a virtual device, this memory access will trap into the hypervisor. 
If the instruction causing the trap takes the value to write from a register, the hypervisor which is
attempting to emulate the access,
will not be able to read it when the general purpose registers are encrypted. 
The \texttt{vmcb} must be extended to contain decode assists for these events to provide the required information.
As indicated in~\cite{lkmlGPREGS} decode assists are already in place to allow the hypervisor to read the instruction causing a \texttt{vmexit}.
For future versions of SEV, these assists must be extended to contain the register values that hold the arguments to the instruction.
The system must only augment \texttt{vmexit} events caused by traps to shared pages with these decode assists,
in order to ensure that a malicious hypervisor cannot force a guest to reveal register content through decode assists. 

\paragraph{Access to the vmcb} 
Usually, the \texttt{vmcb} is configured only once during the initial setup while at runtime a benign hypervisor does not need to modify the vmcb, with some exceptions. 
The fact that SEV allows us to alter the \texttt{vmcb} nevertheless imposes a security risk as it permits us to divert the control flow of guest by setting an arbitrary instruction pointer.
We propose to change the existing state caching mechanism to enable the creation of a write-once \texttt{vmcb}.
Currently, the content of the \texttt{vmcb} is already cached to improve context switch performance. 
    The CPU is permitted to use the cached values of the \texttt{vmcb} unless the hypervisor explicitly clears bits in a special \texttt{vmcb} area called \textit{vmcb clean field} and thereby forces the CPU to reread \texttt{vmcb} data.
By prohibiting the hypervisor from altering this field, the CPU is always able to use the cached values.
At the first start of a guest the CPU copies the hypervisor provided \texttt{vmcb} into the cache. 
During runtime, the system always uses the cached \texttt{vmcb}.
The initial \texttt{vmcb} can be assumed trustworthy because it is taken into account for the remote attestation.
If the hypervisor wants to schedule another guest, hence another \texttt{vmcb} must be loaded, the system must provide a way to store the cached \texttt{vmcb} encrypted in shared memory.

However, there are elements in the \texttt{vmcb} which the hypervisor
must be able to modify at runtime. Most importantly, injecting
virtual interrupts into a VM requires the modification
of several fields, among them \texttt{V\_IRQ}, \texttt{V\_INTR\_PRIO},
\texttt{V\_INTR\_MASKING} and \texttt{V\_INTR\_VECTOR}. Efficient
injection of multiple pending interrupts also requires access to the
\texttt{EVENTINJ} field and the \texttt{VINTR} bit in the
generic instruction intercept selection bitmask.
These elements would either have to be excluded from our proposed
``mandatory caching'' scheme, or AMD's interrupt controller virtualization
(AVIC) could be declared as a dependency of SEV thus making those \texttt{vmcb} elements obsolete.

\paragraph{Access to guest memory}
Writes by the untrusted hypervisor to guest memory are dangerous.
The fact that no memory authentication is in use opens the door for fault injection and replay attacks as presented in this paper.
The most common way to protect memory from unauthorized access are integrity trees.
However, they induce a notable performance and memory space overhead \cite{rogers2007}.
In a more relaxed attack model where physical attacks such as bus intercepts or direct memory accesses are not considered, 
it is sufficient to prevent the hypervisor from writing encrypted guest memory using mechanisms such as \textit{CIP} as presented in~\cite{szefer2012architectural}.
The exclusion of pages from hypervisor access requires nontrivial changes to the guest operation system as well as the hypervisor.
Further, the proposed access restrictions impact or even prohibit major cloud maintenance operations like snapshotting or live migration.
Intel's SGX technology uses both encryption and integrity checks to protect the memory of enclaves~\cite{sgx2016}.
However, SGX enclaves are small compared to VMs, and it is thus 
still an open question whether protecting the memory of complete VMs by integrity trees is feasible.

\section{Related Work}
\label{sec:related_work}
In the following section, we present some topics which are relevant to this work.

\subsection{Attacks}
While attacks against AMD's SEV have not been published, several attacks against similar systems have been proposed.
Checkoway et al. \cite{Checkoway2013} proposed an  attack method dubbed Iago 
whereby a malicious kernel manipulates system call return values to mount 
arbitrary code execution attacks on a system that protects userland applications from a malicious kernel.
This work clearly shows that it is important to secure the system call interface from an adversary.
Linux system calls can be identified by a unique number that is stored in the general purpose registers.
As these registers are still subject to manipulation by a hypervisor, this type of attack is also applicable to AMD SEV.

Xu et al. \cite{Xu2015} showed how secret data can be extracted by inferring from page faults that specific execution paths inside a protected SGX enclave were executed.
Using these execution traces, they were able to reconstruct images that were processed inside this enclave.
Their approach of inferring memory content based on pagetable fault information
is similar to the approach used in the proposed replay attack.
SGX however does not hide the process internal address mapping from the attacker, 
which allows for a much more direct method of inference.
Further, they did not deal with multiple concurrent processes.

Weichbrodt et al. proposed an attack dubbed AsyncShock \cite{Weichbrodt}.
They exploit the fact that the operating system is responsible for scheduling SGX enclave threads. 
By forcing enclave exits during the execution of multithreaded enclave code, 
they were able to mount use-after-free and TOCTTOU attacks on SGX protected enclaves.

Similar to our replay attack, Branco et al. \cite{branco2016blinded} exploit the lack of memory authentication
to compromise an encrypted system.
The compromise is accomplished by injecting faults into critical state areas of the login process via the JTAG interface.

\subsection{Defenses}
Protecting applications from higher privileged software has been the subject of research for a long time. 
Many solutions that target single applications were proposed such as \cite{chen2008overshadow, hofmann2013inktag, cheng2013appshield, mccune2008flicker}.
Many of these solutions assume the existence of a trusted hypervisor to enforce protection of single applications or parts of an application.

A different direction is explored in the publications
\cite{zhang2011cloudvisor, jin2011architectural, szefer2012architectural, szefer2012architectural}.
The goal of their research is to provide protection mechanisms that ensure the integrity and confidentiality of the guest even in the case of a compromised hypervisor.
Zhang et al. proposed CloudVisor~\cite{zhang2011cloudvisor}
where a trusted security manager provides protection of guest VMs 
by means of nested virtualization.
In contrast, Seongwook et al. proposed \textit{H-SVM}~\cite{jin2011architectural}, a purely hardware-based mechanism to protect guest systems.
The guest memory is not mapped into the hypervisor context and a new hardware component, \textit{H-SVM}, is controlling the nested pagetable. 
The hypervisor cannot access guest memory as it cannot create mappings itself, because the nested pagetables are protected. 
\textit{H-SVM} protects the guest state by setting aside a dedicated memory area that is also not accessible by the hypervisor. If the hypervisor needs to access guest memory, the corresponding page must be explicitly marked by the guest.
Physical attacks are not considered by \textit{H-SVM}.

Similarly, Szefer et al. presented HyperWall~\cite{szefer2012architectural}.
Instead of removing the hypervisor's ability to manage the nested pagetable, an additional protection mechanism is introduced: \textit{Confidentiality and Integrity Protection tables}, short \textit{CIP}.
These tables are consulted by the MMU when accessing memory.

Xia et al \cite{xia2013architecture} followed this path with \textit{HyperCoffer} and added protection against physical attacks by using encrypted memory with integrity checks. 
In this later publication they also address the lack of support for common cloud maintenance operations,
like live migration or VM snapshotting and restoration.

\section{Future Work}
\label{sec:future_work}
After the initial publication of this paper,
AMD released a new version of their Programmer's Manual~\cite{amd64architecture}.
The updated version details a new set of features called SEV-ES,
which encrypts the guest state and enables the guest to finely control which state to share with the hypervisor.
Therefore the first two attacks described in \ref{sec:introduction} are only effective against systems without SEV-ES support.
Further research is needed to examine which attack vectors, besides replay, remain despite the proposed mechanisms. 

While we clearly show, that SEV and SEV-ES cannot offer protection against a malicious hypervisor,
we are confident that those technologies will thwart a substantial amount of attacks that rely on fewer capabilities.
The question to what extend less severe hypervisor bugs, 
which do not lead to a complete compromise, but rather cause data leakage or allow to rewrite the memory of another VM
impact the confidentiality and integrity of the tenant's data, has yet to be examined.
%

\section{Conclusion}
\label{sec:conclusion}

This paper presents a first security evaluation of the upcoming Secure Encrypted Virtualization technology by AMD.
While there are no actual CPUs available yet, the official documents 
published by AMD give away design issues that can be exploited by a malicious hypervisor.\\
By implementing three proof-of-concept attacks, we
showed that these issues can be exploited to fully circumvent the protection mechanisms introduced by SEV.
Furthermore, we showed that even when the hypervisor is not able to control the guest 
using the \textit{vmcb} and general purpose registers,
the control over the nested pagetable combined with the ability to inject interrupts is enough to mount a replay attack.
We proposed possible hardware extensions to mitigate our attacks and compared
similar solutions presented by the scientific community.
Although we discovered serious design issues of AMD's SEV,
we still think that the technology is promising considering the mitigations discussed in this paper.

\bibliographystyle{abbrvnat}
\bibliography{references}

\end{document}